\documentclass[conference]{IEEEtran}
\IEEEoverridecommandlockouts
\usepackage{cite}
\usepackage{amsmath,amssymb,amsfonts}
\usepackage{algorithmic}
\usepackage{algorithm}
\usepackage{graphicx}
\usepackage[font=footnotesize,caption=false,subrefformat=parens,labelformat=parens]{subfig}
\usepackage{textcomp}
\usepackage{xcolor}

\DeclareMathOperator*{\argmax}{arg\,max}

\def\BibTeX{{\rm B\kern-.05em{\sc i\kern-.025em b}\kern-.08em
    T\kern-.1667em\lower.7ex\hbox{E}\kern-.125emX}}
\usepackage{eso-pic}
\newcommand\AtPageUpperMyright[1]{\AtPageUpperLeft{%
 \put(\LenToUnit{0.5\paperwidth},\LenToUnit{-1cm}){%
     \parbox{0.5\textwidth}{\raggedleft\fontsize{9}{11}\selectfont #1}}%
 }}%
\newcommand{\conf}[1]{%
\AddToShipoutPictureBG*{%
\AtPageUpperMyright{#1}
}
}

\begin{document}

\title{UAV Aided Search and Rescue Operation Using Reinforcement Learning
\thanks{This work is supported by NSF under the awards CNS-1453678 and CNS-1916766.}
}
\author{\IEEEauthorblockN{Shriyanti Kulkarni,~Vedashree Chaphekar,~Md Moin Uddin Chowdhury,~Fatih Erden, and Ismail Guvenc}
\IEEEauthorblockN{\textit{Dept. Electrical \& Computer Eng., NC State University}, Raleigh, NC}
\IEEEauthorblockN{Email: \{spkulka2, vschaphe, mchowdh, ferden, iguvenc\}@ncsu.edu}}
\IEEEoverridecommandlockouts
\IEEEpubid{\makebox[\columnwidth]{978-1-7281-6861-6/20/\$31.00~\copyright2020 IEEE \hfill} \hspace{\columnsep}\makebox[\columnwidth]{ }}
\maketitle
\conf{IEEE SoutheastCon2020}
\pagestyle{plain}
\begin{abstract}
Owing to the enhanced flexibility in deployment and decreasing costs of manufacturing, the demand for unmanned aerial vehicles (UAVs) is expected to soar in the upcoming years. In this paper, we explore a UAV aided search and rescue~(SAR) operation in indoor environments, where the GPS signals might not be reliable. We consider a SAR scenario where the UAV tries to locate a victim trapped in an indoor environment by sensing the RF signals emitted from a smart device owned by the victim. 
% We assume that the smart device periodically transmits RF signals to nearby wireless access points. 
To locate the victim as fast as possible, we leverage tools from reinforcement learning~(RL). 
% where we consider that the UAV and the victim's device are equipped with directional antennas. 
Received signal strength~(RSS) at the UAV depends on the distance from the source, indoor shadowing and fading parameters, and antenna radiation pattern of the receiver mounted on the UAV. To make our analysis more realistic, we model two indoor scenarios with different dimensions using a commercial ray tracing software. Then, the corresponding RSS values at each possible discrete UAV location are extracted and used in a Q-learning framework. Unlike the traditional location-based navigation approach that exploits GPS coordinates, our method uses the RSS to define the states and rewards of the RL algorithm. We compare the performance of the proposed method where directional and omnidirectional antennas are used. The results reveal that the use of directional antennas provides faster convergence rates than the omnidirectional antennas.
\end{abstract}


\section{Introduction}
\label{sec:introduction}
Due to their ease of deployment, high flexibility, and low manufacturing cost, drones or unmanned aerial vehicles (UAVs) can be deployed in various civilian applications, including but not limited to precision agriculture, cleaning up ocean waste, packet delivery, restoring wireless service after natural disasters, and search and rescue (SAR) operations~\cite{mozaffari_uav,8468611,8714567,chowdhury2019rssbased,ezuma2019microuav}. It  is  estimated  that  the  market  for  commercial UAVs  will  reach  a multi billion U.S. dollars  by  the  year 2025~\cite{estimate}. Hence, a huge amount of efforts in both academia and industry have been devoted to different aspects of UAV applications.

%The use of unmanned aerial vehicles (UAVs) is a promising technology for next-generation cellular networks. It is estimated that the market for commercial UAVs will reach a multi billion U.S. dollars  by  the  year  2025\cite{estimate}. 

%The growing importance of wireless communications, internet and networking has led humans towards Industry 4.0. Extensive work in wireless communication has resulted in the evolution of LTE and 5G. The unmanned aerial vehicles (UAVs) is proving to be a low-cost solution enabling high mobility which finds great importance in the telecommunication industry. These are agile and provide higher coverage area which lead to quicker results in search and rescue operations. UAVs find use not only for tackling problems like disease control, cleaning up ocean waste, delivering pizza, but also help to identify inaccessible targets \cite{chowdhury2019rssbased}, \cite{8468611}, \cite{8714567}. 

% \blfootnote {This work is supported by NSF under the awards CNS-1453678 and CNS-1916766.}

SAR operations occur during or after natural disasters, finding missing persons, accidents, evacuation process to provide immediate help. These operations are time-critical and proven to be effective when the operation is completed in the shortest possible time. These operations can take place where human access may not be possible. Thanks to their flexibility in deploying, UAVs are suitable for such scenarios and hence, can aid interaction to inaccessible or unapproachable environments for humans~\cite{autonomous}. However, due to the lack of a reliable communication link between the UAV and the controller or technical requirements beyond human capabilities, human control over the UAVs may not always be possible~\cite{Becerra}. Hence, it is critically important to develop effective yet accurate algorithms to enable the UAVs to perform complicated tasks without human interaction. Note that in the case of SAR operations, most of the time, the prior knowledge pertinent to the environment is limited, if not completely unavailable. Furthermore, the environment in interest can change abruptly or the models defining the victim's locations may not be accurate. Hence, the autonomous UAV deployed in the SAR operation can interact with the environment for taking decisions by itself. In such a case, a subclass of machine learning called \textit{reinforcement learning} (RL), which does not require any prior knowledge of the environment, may help to overcome these issues.

\begin{figure}
\centering{\includegraphics[trim=2cm 16.5cm 3cm 3.5cm, clip,width=\linewidth]{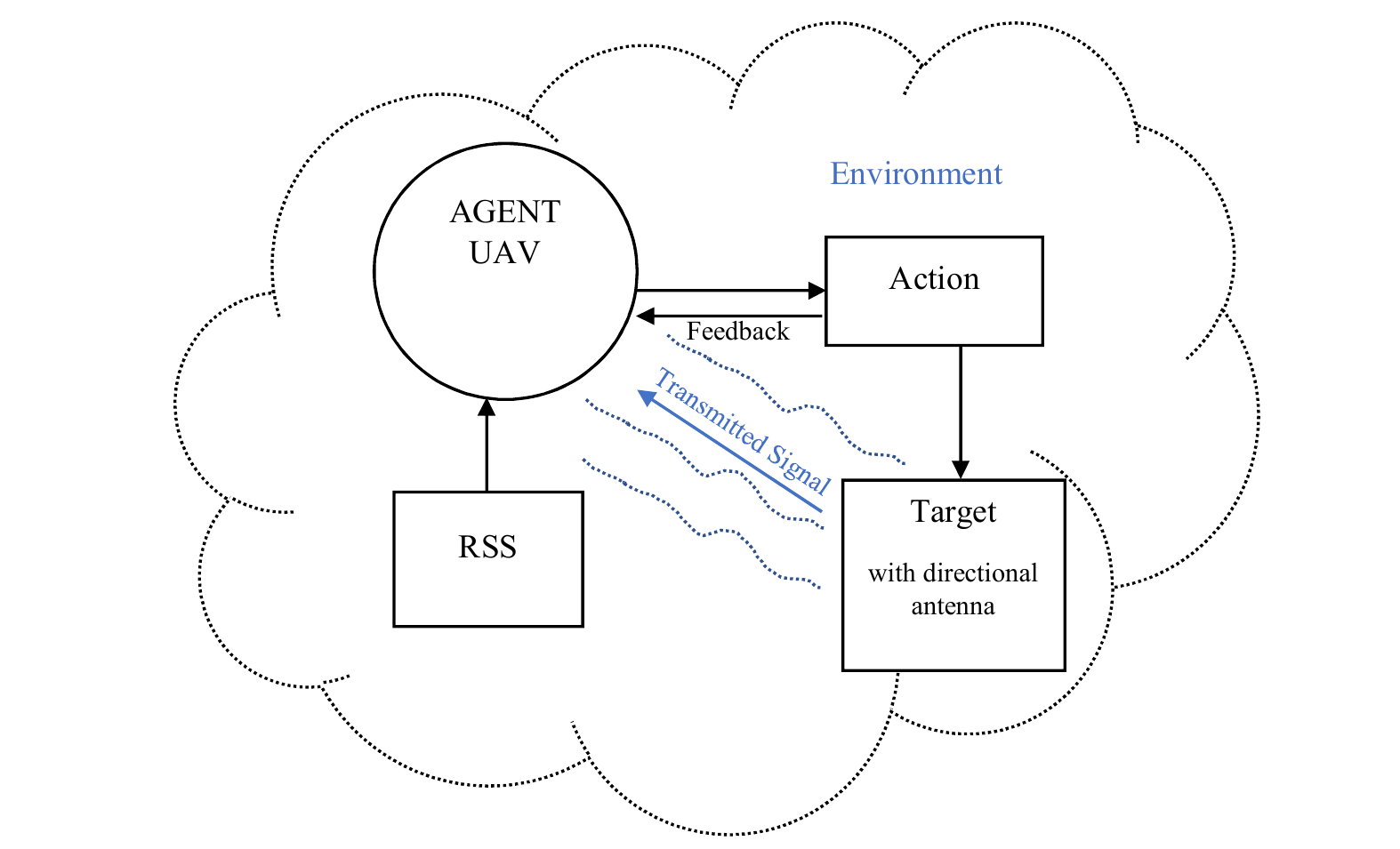}} 
\caption{Proposed system design for SAR operation using a directional antenna.}
\label{fig:prposed_sytem}
\end{figure}

\IEEEpubidadjcol
In a typical RL algorithm, the agent learns how to act properly in an unknown environment by repetitive interaction with that particular environment~\cite{Sutton1998}. The environment is usually modeled as a  Markov decision process (MDP), and a dynamic programming (DP) technique is used to find the optimum set of sequential actions or the optimum \textit{policy}~\cite{dp}. However, to utilize the DP technique, the complete knowledge of the environment is required to be known beforehand. On the contrary, a branch of RL, known as Q-learning, does not require such explicit knowledge of the environment. Q-learning is an off-policy RL algorithm which aims to find the best policy for maximizing total discounted future reward in an unknown environment given the current state~\cite{Sutton1998}.

%The typical model involving an inanimate rescuer is made up of an agent and the environment. Here, the UAV is the agent and the target is the RF device placed in an environment unknown to the agent. To help the UAV interact with the environment, we made use of machine learning which helped the device gain knowledge about its environment. For this purpose, a subclass of machine learning called reinforcement learning (RL) was used, which does not need any prior knowledge of the environment. 

RL-based algorithms have already been studied in UAV-related researches such as robotics, wireless communication, etc~\cite{gesbert2018, Iman2016}. The authors in~\cite{gesbert2018} explore both traditional table-based and neural network-based Q-learning to provide better connectivity in a downlink cellular network. In~\cite{Iman2016}, a model-based RL algorithm is used for the autonomous navigation of UAVs, which learns faster than the traditional table-based Q-learning due to its parallel architecture. In~\cite{8468611}, a function approximation-based RL approach is exploited for a large number of states. The use of function approximation reduces the convergence time of the algorithm needed for a SAR operation. This approach also considers the obstacle avoidance needed by an agent to avoid a collision in the environment. The results were proved using simulation as well as real-time implementation. In~\cite{Sampedro2019}, the authors implement a SAR operation for a fully autonomous robotic solution in unstructured indoor environments. The target recognition uses a supervised learning approach for target background classification. 
% In \cite{ezuma2019microuav}, authors use a Naive Bayes approach which was based on Markov models to detect and classify micro-UAV control signals. It is based on the generation of normalized energy trajectory which is obtained from the energy-time-frequency distribution of raw control signals. It uses neighborhood component analysis to extract significant features that are detected from abrupt changes in the mean of energy trajectory. This ensures that the computational cost of the algorithm is lower. 
The study in~\cite{g2017learning} uses deep deterministic policy gradients to provide results in continuous time and action. It also uses a framework that makes sure that the UAV uses a collision-free path to reach the target. The UAV not only focuses on rewards, but also considers the punishment irrespective of the action taken. The use of neural networks and deep learning was briefly performed in~\cite{Lygouras_2019}. This is an embedded system-based approach that guarantees safe take-off, navigation, and landing of the UAV on a fixed target. .\looseness=-1

In a recent study, the authors propose a SAR method by exploiting the RF signal emitted from the devices owned by a victim~\cite{chowdhury2019rssbased}. They also provide a comparison between the location-based and received signal strength~(RSS)-based approach using an omnidirectional antenna for an indoor scenario. In this approach, the target’s location, as well as the environment, are unknown to the agent, whereas for the location-based approach, the target position and the UAV's instantaneous location are known based on which the agent rescues the target. The comparison also proves that the learning phase of the RSS-based approach has lesser fluctuations in training the agent than that in the location-based technique. 
% Note that authors consider omnidirectional antennas both at the transmitter owned by the victim and the receiver mounted on the UAV~\cite{chowdhury2019rssbased} and there was a limitation of convergence in time. 
Since the directional antenna has greater gain in a concentrated direction, it can be more effective to find targets in areas with weak signal power. The directional antenna also provides a larger coverage area, better isolation, and lesser feedback loops. The direction in which the UAV receives the most powerful signal helps understand the probable location of the target. Note that the GPS signals in an indoor environment might not be reliable and hence, the RSS-based indoor navigation is a competitive candidate for SAR missions in indoor scenarios as shown by the authors in~\cite{chowdhury2019rssbased}.

Motivated by the above references, we propose a Q-learning based SAR operation method with the presence of directional antennas both at the UAV and the victim's device by sensing the RSS emitted from the victim's device. In case of emergencies, UAV can also force the smart-device to transmit wireless signals~\cite{Pu2019,wang}. It is also of critical importance to ensure that the UAV reaches the target as soon as possible without colliding with any of the obstacles present in the environment. The UAV in interest senses the RF signals which are emitted intermittently from the victim to navigate through the indoor environment. Prior knowledge of the environment and the location of the victim are not known, and no training set is available for the UAV to map a particular action set to the reward or punishment. The RSS values at different indoor locations can be used by the UAV to take actions to reach the target device without having prior knowledge of the environment.  Initially, there is no presence of past experiences or rewards and punishments based on which the UAV can take actions. The Q-learning algorithm uses a Q-table and includes flags for punishment and reward. These flags were updated on every move, which, in turn, generate a Q-table for the UAV from its own actions. As the Q-table gets updated at each iteration, the probability of taking a good action will improve, leading to a faster approach to the target device.

  %Note that since Thus, the UAV has information of both, received power as well as physical obstructions in the environment aiding it to reach the target. 

%In this paper, we consider Q-learning which is an off-policy RL algorithm. Based on its current location, the agent takes an action to reach the target device. The agent receives corresponding feedback in the form of reward or punishment which classifies the action taken as productive or unproductive in the search and rescue operation. The consequence of each action is noted in a table by the agent for future reference. The agent keeps making random actions which were not previously known until it reaches a policy that maximizes the reward \cite{chowdhury2019rssbased}.  

% \begin{figure}
% \centering{\includegraphics[trim=0.7cm 18cm 1cm 4cm, clip,width=\linewidth]{f12_v2.pdf}} 
% \caption{Proposed system design for SAR operation using a directional antenna.}
% \label{fig:prposed_sytem}
% \end{figure}

\begin{figure}[t]
\centering{\includegraphics[trim=0.7cm 18cm 1cm 4cm, clip,width=\linewidth]{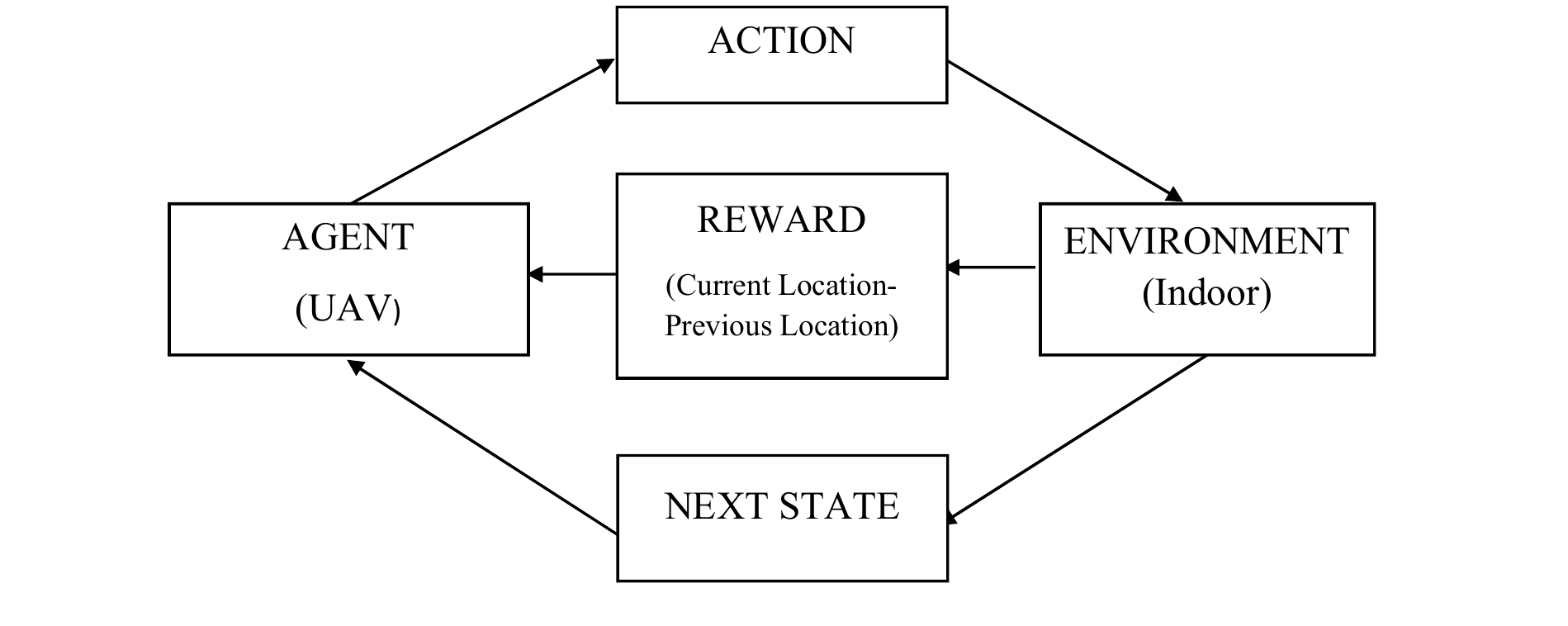}} 
\caption{Interaction between agent and environment in Q-learning.}
\label{fig:interact_q_learning}
\end{figure}

%The UAV then takes necessary actions to try to reach to the most proximate target device.This acquisition is to be made in the least possible distance and time. SAR operation is a time dependent operation. Lesser the time the UAV takes to find and reach the target, faster is the target rescued. For the agent to reach the target a set of 8 actions were devised. Using either of these actions for every state, the UAV traverses the environment to reach the next highest received power location to eventually reach the target.For this realization, Q-learning approach was used. The UAV was designed to pick the best action which gave the highest valued reward to reach the target. Rescuing the target being the highest reward, the agent acts greedily and keeps taking actions to attain this reward. The UAV keeps its search on-going to evacuate all the target devices. Reinforcement Learning is used for this approach.  This approach helped in searching and rescuing the device. In inaccessible areas or GPS restricted locations, humans have no prior knowledge about the setting of the location. As, the UAV collects rewards and punishments from its own actions, the size of the Q-table is a crucial factor in determining the time taken by the UAV to rescue the target.

\begin{figure*}[t]
\centering
\subfloat[]{
\includegraphics[width=.35\linewidth]{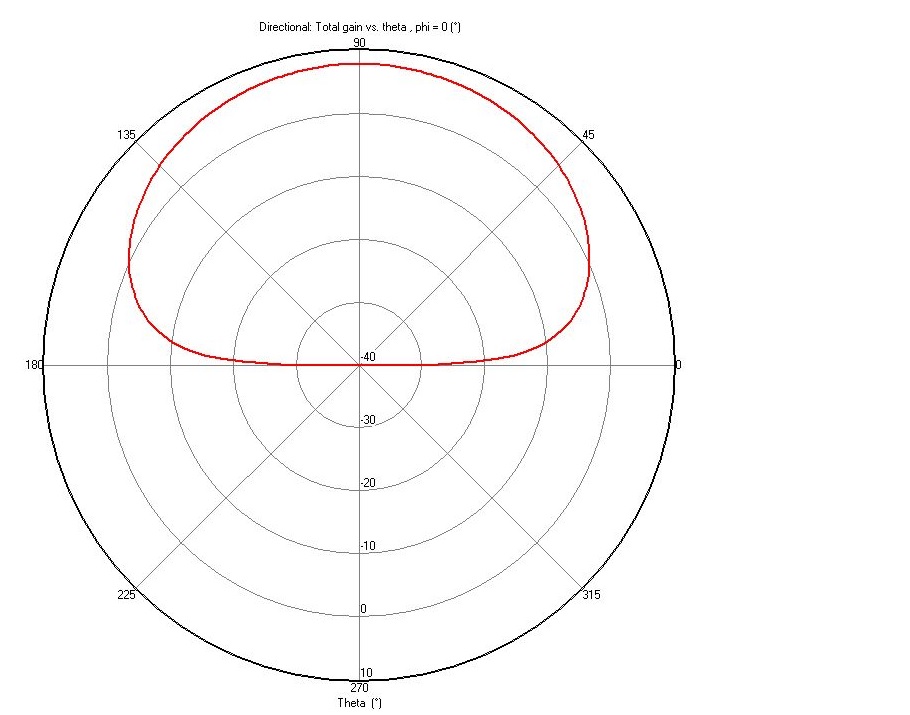}}
\subfloat[]{
\includegraphics[width=.35\linewidth]{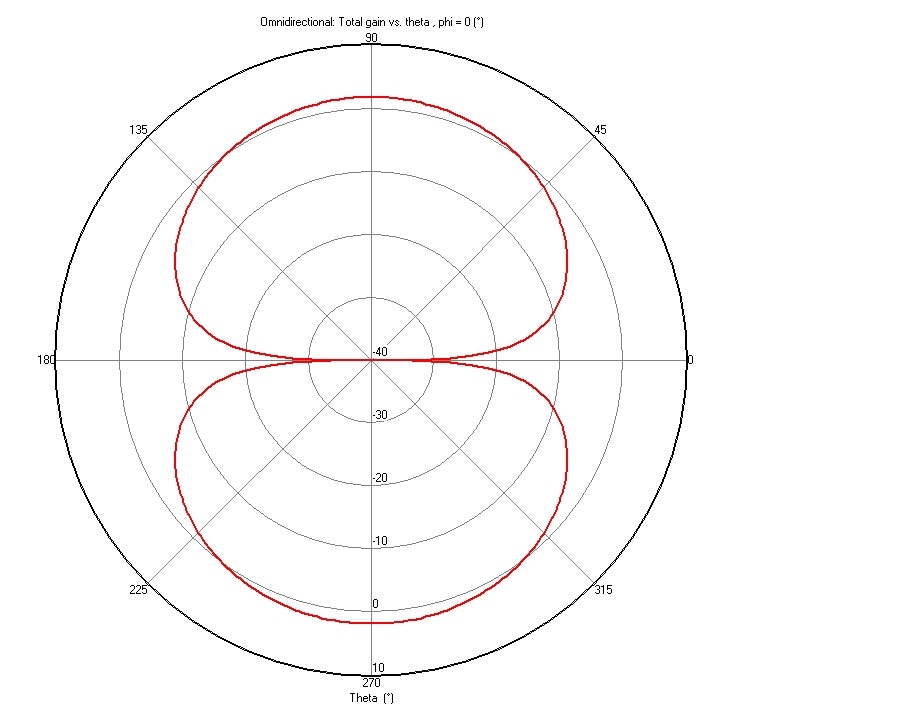}}
\caption{Radiation patterns for (a) directional and (b) omni-directional antennas.}
\label{fig:antenna_radiation}
\end{figure*}  

A high-level view of our proposed SAR operation is depicted in Fig.~\ref{fig:prposed_sytem}. The agent or the UAV takes actions in an indoor environment depending on the RSS, which is obtained from the RF signals transmitted by the victim. A unique state label is assigned to the RSS value at the current UAV position as done in~\cite{chowdhury2019rssbased}. Based on its current state, the UAV chooses an action in different directions separated by $45^{\circ}$ to reach the next location, and this process goes on until the UAV finds the victim. The result of the action taken by the agent is given in the form of feedback. This feedback is either a reward or punishment. 
% We discuss the state and reward definition process in Section \ref{sec:indoor_uav_navigation}. 
A ray tracing software is used to create two different indoor environments that involve obstacles in the form of doorways, windows, and indoor walls. After simulating in the ray tracing software, the obtained RSS values are used as an input to generate a heat map in MATLAB, which is used to define both the states and the rewards. The proposed method was compared to the results obtained using an omnidirectional antenna in terms of the convergence time and the number of steps taken to reach the victim.

%Our contribution in this paper can be summerized as follows:
The remainder of this paper is organized as follows. Section II provides a brief background on Q-learning while Section III discusses the antenna radiation patterns associated with both directional and omnidirectional antennas. Section IV introduces the simulation setup. Section V gives a brief description of the navigation of the UAV in the indoor environments used. The simulation results of the proposed SAR operation are presented in Section VI, and finally, the paper is concluded in Section VII.

\section{Background on Q-Learning}
As stated earlier, RL involves interaction between an active decision-making agent and its environment, within which the agent seeks to maximize the future reward despite the uncertainty associated with the environment. An RL agent interacts with the environment in the form of discrete time steps. At each time step, the actions taken by the agent determine the future states, thereby affecting the options and opportunities available to the agent at later times. Correct choice requires considering indirect, delayed consequences of actions, and thus may require foresight or planning. The training phase allows the agent to gain this foresight that helps the agent attain its final goal efficiently with high accuracy. Through repeated action selections, the agent can maximize winnings by concentrating the actions on the best value which is obtained from the options available at a particular time step. If the agent maintains the estimates of the action values, then at any time step there is at least one action whose estimated value is the greatest. Fig.~\ref{fig:interact_q_learning} illustrates an example of the interaction between the agent and the environment.

If a greedy action is selected, it is said that the agent is exploiting current knowledge of the values of the actions. If instead the agent selects one of the non greedy actions, then we say it is exploring, because this enables the agent to improve the estimate of the non greedy action’s value. On one hand, exploitation is the right thing to do to maximize the expected reward, while on the other hand, exploration may produce greater total reward in the long run. Thus, using the greedy policy will not allow the agent to explore at all. In order to still allow some exploration, an $\epsilon $-greedy policy is used. This involves selection of an action (or one of the actions) with highest estimated action value, that is, to select one of the greedy actions at some time step $t$. Value at the current state in the Q-table is calculated as reward plus the value at the next state. This value is updated in the Q-table. Q-learning is a temporal difference method in which we measure all outcomes at the next state to get an estimate of the value at the current state. It learns by interacting with the environment and approximates a value function of each state-action pair through a number of iterations. The goal is to select the action which has the maximum $Q$-value using the following update rule at~each~iteration:
\begin{equation}
\label{eq:Q_Update_eqn}
{Q(s,a)\leftarrow(1-\alpha)Q(s,a)+\alpha\big[r(s)+\gamma\max\limits_{a'}Q(s',a')\big]},
\end{equation}
where $a^\prime$ is the action taken at time $t+1$, $s^\prime$ represents the state reached from state $s$ after action $a$ is taken, $\alpha \in [0,1] $ is the learning rate, $r(s)$ is the reward attained at current state $s$ and $\gamma \in [0,1)$ is the discount factor. After the iterative process, the agent will eventually learn the optimal Q-values for each state-action pair, $Q^*(\mathbf{s},\mathbf{a})$ over time. The actions with the highest Q-values for
each state represent the optimal policy.  In other words, the optimal policy can be obtained by acting greedily in every state by
\begin{equation}
\label{eq:opt_policy}
\pi^*=\argmax\limits_\mathbf{a} Q^*(\mathbf{s},\mathbf{a}).
\end{equation}

% Here, equation \eqref{eq:opt_policy} denotes the value of $a$ at which the equation is maximized \cite{Sutton1998}. 

\begin{figure*}[t]
\centering
\subfloat[]{
\includegraphics[width=.39\linewidth]{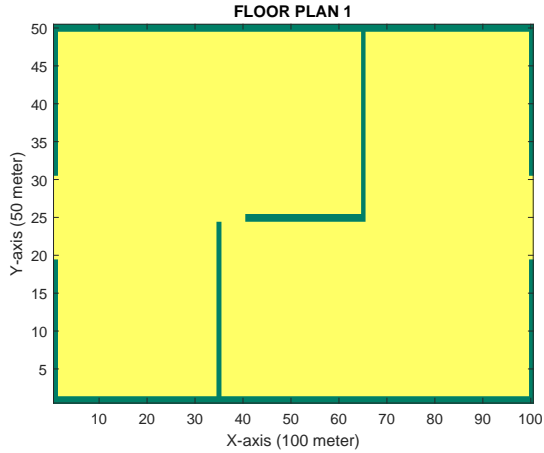}}
\hspace{2 cm}
\subfloat[]{
\includegraphics[width=.42\linewidth]{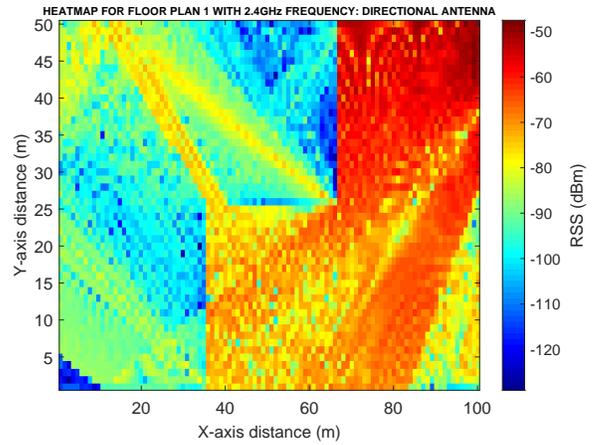}}
\caption{(a) Floor plan of the indoor environment and (b) the corresponding RSS heat map.}
\label{fig:floor_plan1}
\end{figure*}

%Greedy action selection always exploits current knowledge to maximize immediate reward; it spends no time at all sampling apparently inferior actions to see if they might really be better \cite{Sutton1998}. 

%To elaborate this dilemma in our SAR operation, we can say that at a given location, the UAV is exploiting the rewards. If the UAV takes the actions which always lead to a reward, then it approaches closer to the victim. On the other hand, when the agent is in the training phase it will be exploring. As the time to explore increases, it helps the UAV to search for the victim in lesser time. UAV attains the highest reward when it reaches the target. If the agent only runs behind this highest reward, it is possible that it may not take the optimal route possible. While, if we reduce the maximum reward to attain the target, the UAV will keep on exploring the possible ways to reach the target as opposed to rescuing it.\cite{nptel_rl}, \cite{brockman2016openai} \textcolor{red}{May need to reword this paragraph}

\label{sec:Setup}

\section{Background on Antenna Radiation}
In this section, we discuss the antenna radiation patterns for omnidirectional and directional antennas which are used by the UAV to find the RSS at each receiver grid. The radiation pattern is defined as the mathematical function of the radiation properties of an antenna as a function of space coordinates. The radiation pattern also provides insight into characteristics like flux density, field strength, directivity, and polarization. A graph of variation of power density amplitude along a constant radius is a power pattern and a trace of the received electric field, magnitude at a constant radius is the field pattern. 

A radiation lobe is a portion of the radiation pattern within a given area~\cite{Balanis-2012-antenna}. From Fig.~\ref{fig:antenna_radiation}, we can see that the omnidirectional antenna has two main lobes and transmits the power in all directions. On the other hand, for the directional antenna, we see only one main lobe, and the transmitted power is concentrated in a specific direction. Directivity is defined as a ratio of radiation intensity in a given direction to the radiation intensity averaged over all directions~\cite{Balanis-2012-antenna}. In weak signal environments, received signal power is concentrated in specific directions, making directional antenna a better choice for our RSS-based SAR scenario.

\section{Simulation Setup}
In this section, we discuss the simulation environment modeling for testing our proposed method. The simulation environments used in this study are modeled by using the Wireless InSite ray tracing tool. This software can provide RSS values at each point in an indoor scenario. We have generated two arbitrary indoor floor plans with diverse obstacles. Table~\ref{tab:floorplan} provides a brief summary of all the floor plans used along with the presence of obstacles. 

\begin{table}[t]
\caption{Floor Plan Summary.}
\centering
\begin{tabular}{ccc}
\hline
&Floor plan 1& Floor plan 2 \\
\hline
Dimensions&	($100~\text{m}\times50~\text{m}$)&	($63~\text{m}\times43~\text{m}$)\\
Doors	&1&	1 \\
Windows&	2&	0 \\
Walls&	7&	10 \\
Victim position	& (45~m, 90~m)	& (38~m, 27~m) \\
\hline
\end{tabular}
\label{tab:floorplan}
\end{table}

In these floor plans, we have considered walls with a  height of 3~m. After generating the floor plans, we run the ray tracing simulations to obtain the RSS at each receiver (RX) grid with the source being stable at a specified position. The RX grid points are set 1 m apart from each other using the XY grid option in the software. It was assumed that the victim transmitted RF signals using 25~dBm transmit power at 2.4~GHz. Directional antenna was used at the transmitter. The maximum antenna gain was considered as 0~dB for both the receiver grid and transmitter\cite{remcom}. We created the same floor plans in MATLAB and transferred the resulting RSS maps from the ray tracing software to MATLAB for using in the navigation simulations. For simulation purposes, the floor areas were partitioned into grids, and hence the UAV was forced to move from the center of one grid to that of the other. Fig.~\ref{fig:floor_plan1} shows the floor plan and the corresponding heat map.\looseness=-1

\begin{algorithm}[t]\small 
	\caption{Q-learning for indoor UAV navigation.}
    \label{alg:Alg_Q-learning}
	\begin{algorithmic}[1]
	%	\STATE initialize a $Q$ table of row = 10000 and column =8
		\STATE  start from an initial location and obtain associated\\ state of that particular location by sensing the RSS
		\STATE \textbf{repeat} (for each step):
		%\STATE \hspace{0.3cm} start from an initial location and obtain associated\\\hspace{0.3cm} state of that particular location by RSS
	%	\STATE \hspace{0.3cm} \textbf{if} $\epsilon$ $\geq$ $\epsilon_{\rm min}$
	%	\STATE \hspace{0.8cm} $\epsilon$=$\epsilon$ $\times$ $\exp{(-\eta)}$ \textbf{end if}\\
		%\STATE \hspace{0.3cm} \textbf{if} $\alpha$ $\geq$ $\alpha_{\rm min}$
		%\STATE \hspace{0.8cm} $\alpha$=$\alpha$ $\times$ $\exp{(-\eta)}$ \textbf{end if}\\
	%	\STATE \hspace{0.3cm} obtain $r(s)$ and $s$ according to the averaged RSS
		\STATE \hspace{0.3cm} choose $a$ using $\epsilon$-greedy policy
		\STATE \hspace{0.3cm} take action $a$, observe $s'$
		\STATE \hspace{0.3cm} check $s'$ for possible obstacle(s)
		\STATE \hspace{0.3cm} \textbf{while} any obstacle at $s'$ \textbf{do}
		\STATE \hspace{0.8cm} leave $a$ and select any other action randomly, \textbf{end while}
		%\STATE \hspace{0.3cm} \textbf{end}
		%\STATE \hspace{0.4cm} $Q(s,a)\leftarrow Q(s,a)+\alpha[r(s,a)+\gamma \max\limits_{a^{'}}Q(s^{'},a^{'})-Q(s,a)]$
		\STATE \hspace{0.3cm} Calculate reward for taking action $a$ by  subtracting RSS\\\hspace{0.3cm}  associated with state $s$ from state $s'$
		\STATE \hspace{0.3cm} update $Q$-value using \eqref{eq:Q_Update_eqn}
        
%        \STATE \hspace{0.3cm} \textbf{end}
		\STATE \hspace{0.3cm} $s \leftarrow s'$
		\STATE \textbf{until} $s$ is terminal
	\end{algorithmic}
\end{algorithm}

In order for the UAV to traverse in the indoor environment, eight actions are predefined which are separated by $45^\circ$ from one another in the $xy$ plane. These actions are north, south, east, west, northeast, northwest, southeast, southwest respectively. \par For the simulation purpose, the floor plan is partitioned into RX grids which are spaced 1~m apart from each other, and the UAV moves from the center of one RX grid to that of the other. The velocity of the UAV is considered to be $v$~m/s. Therefore, as the actions are separated by an angle of $45^\circ$, when the UAV makes a movement in a diagonal direction, its velocity will be $v\sqrt{2}$~m/s. For simplicity, while presenting the results in Section~\ref{sec:results}, we will refer to the UAV speed as $v$~m/s independent of the movement direction. We also assume that the UAV senses the RSS in every 1~s. In other words, the UAV will detect the RSS values only when it reaches a new location from the previous location after performing an action. The overall Q-learning process is presented in Algorithm~\ref{alg:Alg_Q-learning}.\looseness=-1

\section{Indoor UAV Navigation}
\label{sec:indoor_uav_navigation}
For the UAV to traverse the indoor environment, its initial position is predefined. Out of the eight possible actions, the UAV will try to select the best actions which will lead it to the victim as soon as possible. To find the balance between the inherent exploration-exploitation dilemma associated with RL, UAV will use the $\epsilon $-greedy approach with an aim to find the optimum policy $\pi^*$. 
% Here we 
\subsection{State Definition}
As stated above, the UAV will start from an initial position and sense the RSS value associated with that position. A state label will be allocated to that particular RSS value. Since any two RX grids separated by 1~m will have different RSS value, each location in the RX grid will be
represented by a unique state. Afterward, the UAV will take action based on the $\epsilon $-greedy approach and move on to the next state. It is worth noting that, each episode or iteration will be ended when the Euclidean distance between the UAV and the victim is less than 2~m. Since the UAV is functioning in a GPS-denied indoor environment, we translate this distance threshold to an RSS value of -21~dBm. In other words, if the RSS value at a position is less than this threshold, we assume that the Euclidean distance between the UAV and the victim is less than 2~m. We calculate this threshold using the free-space path loss model~\cite{fspl}.
%\par Every action taken by the UAV leads to a value and ience stored by the UAV in the Q-table.The size of the Q-table is dictated by the size of the indoor environment and the eight predefined actions.  The process of navigation of the UAV yields all the values of table which enables it to reach the victim. Every state in the Q-table is uniquely defined with respect to the action taken at the corresponding receiver grid position considered in the ray tracing software and its corresponding reward.  
\subsection{Reward Definition}

In our proposed method, the reward function is defined as the difference between the RSS values associated with the two subsequent UAV positions, i.e., $\textrm{RSS}_t-\textrm{RSS}_{t-1}$, so that higher rewards are obtained when there is an increase in the RSS. Since the victim is transmitting the RF signal in the environment, such a reward function will encourage the UAV to traverse through the locations with higher RSS values and subsequently locate the victim. 
%\par The feedback mechanism used in RL is in the form of rewards and punishments. Rewards are calculated by taking the difference between the current position and the previous position. A reward is given when the UAV takes an action which is beneficial for the SAR operation. That is, the UAV gets closer to the target or takes a step towards the victim without collision. A punishment is given to the UAV when the action taken is away from the victim or not advantageous for the SAR operation. In order to avoid collision in the indoor environment, any action that leads to collision with the features of the indoor environment is given a high negative value. In addition to providing high negative values to bad actions, finding the victim is assigned high positive value. In this SAR operation, the UAV concludes that its search is completed when it reaches the 2~m periphery of the victim. 

%\begin{algorithm}
% \hline
%\caption{Q-learning using $\epsilon$-greedy approach for UAV navigation. \textcolor{red}{This pseudo-code is not clear enough for reproducing}}
% \hline
%\begin{algorithmic}[1]
%\STATE 
%UAV starts from a predefined location (x,y)
%\STATE
%Pick one  action from the predefined 8 actions from this location and use $\epsilon $- greedy algorithm
%\STATE
%Pick an action to estimate the current state reward corresponding to the action
%\STATE
%Update the Q-Table with corresponding reward and the action taken for it.
%\STATE
%Go to the next state
%\STATE
%Repeat for all states.
%\end{algorithmic}
% \hline
%\end{algorithm}

\begin{figure}[t]
\centerline{\includegraphics[width=9cm]{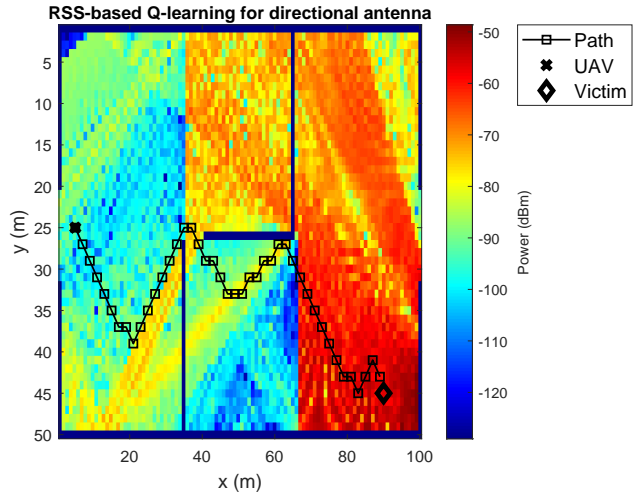}}
\caption{Trajectory followed by UAV to reach victim on floor plan 1.}
\label{fig:floorplan1}
\end{figure}

\section{Simulation Results}
\label{sec:results}

\begin{figure}[t]
\centerline{\includegraphics[width=9cm]{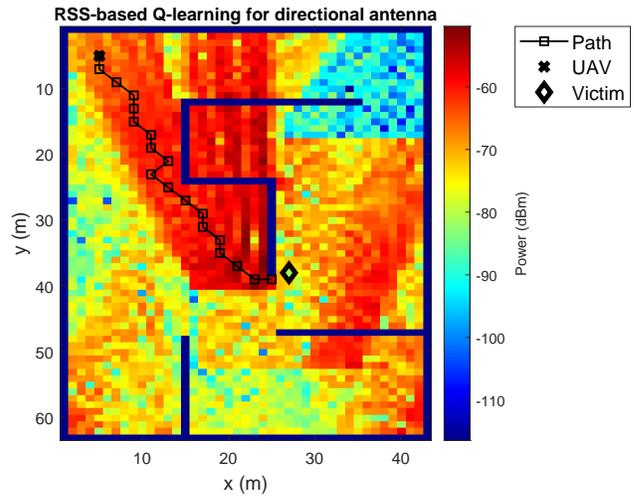}}
\caption{SAR operation done by UAV on floor plan 2.}
\label{fig:floorplan2}
\end{figure}

After realization of the proposed algorithm, we implemented the SAR operation for 30000 iterations on different floor plans using a directional antenna. The details of the floor plans used are mentioned in Table ~\ref{tab:floorplan}. The results of the algorithm implemented on the floor plans are discussed further.  Fig.~\ref{fig:floorplan1} shows the trajectory followed by the UAV to reach the target which is implemented on floor plan 1. Similarly, it can be seen in Fig.~\ref{fig:floorplan2} the trajectory followed by UAV on floor plan 2. Both these floor plans have different features and proved to be effective to analyse the performance of the SAR operation.

\begin{table}[t]

\caption{Results for Floor Plan 1.}

\centering
\begin{tabular}{cc}
\hline
Parameter & Value\\
\hline
UAV start position& (25~m, 5~m) \\
Target location&	(45~m, 90~m)\\
Number of Iterations	&30000 \\
Total time& 74.7~s\\
Trajectory length& 111.34~m \\
Total steps taken	&43 \\
\hline
\end{tabular}
\label{tab:results_floorplan1}
%\end{center}

\end{table}

\begin{table}[t]
\centering
\caption{Results for SAR on floor plan 2.}
%begin{center}
\begin{tabular}{cc}
\hline
Parameter & Value\\
\hline
UAV start position& (5~m, 5~m) \\
Target location&	(38~m, 27~m)\\
Number of iterations	&30000 \\
Total time & 20.84~s \\
Trajectory length& 45.11~m \\
Total steps taken	&19 \\
\hline
\end{tabular}
\label{tab:results_floorplan2}
%\end{center}
\end{table}

The SAR operation can be determined as successful only when the UAV is able to locate and reach the target. In order to ensure that the UAV completed the SAR operation without collision, we had predefined multiple start positions for the UAV in existing indoor obstacles and obtained the following results. 

\begin{figure}[t]
\centerline{\includegraphics[width=9cm]{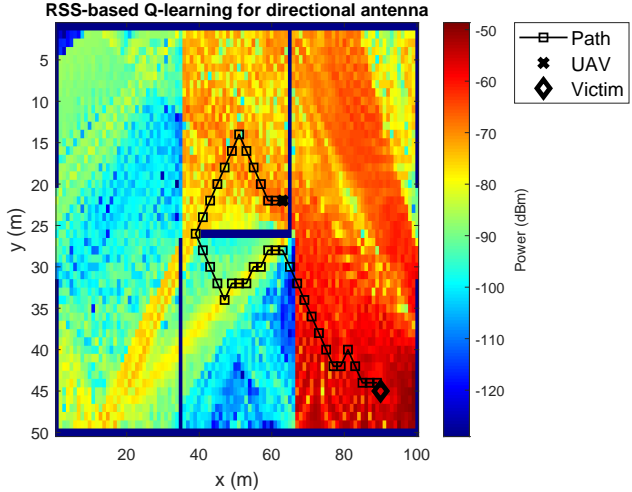}}
\caption{Collision-free trajectory followed by UAV to rescue victim.}
\label{fig:collision}
\end{figure}

From Fig.~\ref{fig:collision}, if the SAR operation was independent of obstacles, then the UAV would have followed a trajectory having the highest signal power which is a straight path. We have also assumed that the width of the indoor walls is greater than 2~m so that if there is a wall in between the UAV and the victim, the distance between them will always be greater than 2~m. From  Fig.~\ref{fig:collision}, we can arrive at the conclusion that even when the UAV had a probable straight path, it followed a different trajectory due to the presence of obstacles.
Next, we have compared the elapsed time, total path length and total number of steps taken by UAV for 1000 and 30000 iterations. The number of iterations indicates the degree of freedom given to the UAV to explore the indoor environment. At every iteration, the UAV updates the Q-table and learns more about the environment. The increase in the number of iterations improves the familiarity of the UAV with the environment as well as the probable actions it can take to reach the target.As the number of iterations increase, the UAV takes some time to train itself from its own experiences however it is done at the cost of more accurate results. Table ~\ref{tab1:1000} and Table ~\ref{tab1:30000} indicate the results obtained on floor plan 1 after altering the number of iterations. As the number of iterations increase, the UAV has more experiences with itself and aids it to reach the target with higher accuracy. 

\begin{figure}[t]
\centering{\includegraphics[width=9cm]{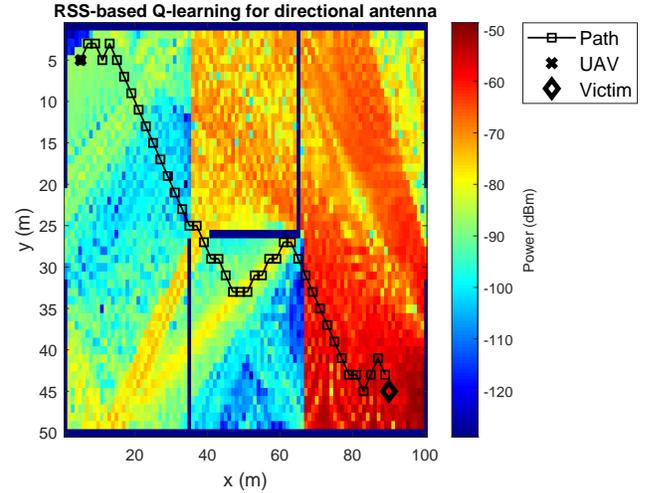}}
\caption{Trajectory of UAV for 1000 iterations.}
\label{fig}
\end{figure}

\begin{table}[t]
\caption{Results for 1000 Iterations.}
\begin{center}
\begin{tabular}{cc}
\hline
Parameter & Value\\
\hline
UAV start position & (5~m,5~m) \\
Target location&	(45~m,90~m)\\
Number of iterations	&1000 \\
Total time&12.56~s\\
Trajectory length&111.34~m \\
Total steps taken	&43 \\
\hline
\end{tabular}
\label{tab1:1000}
\end{center}
\end{table}

\begin{figure}[t]
\label{tab:results_10000_iter}
\centering{\includegraphics[width=9cm]{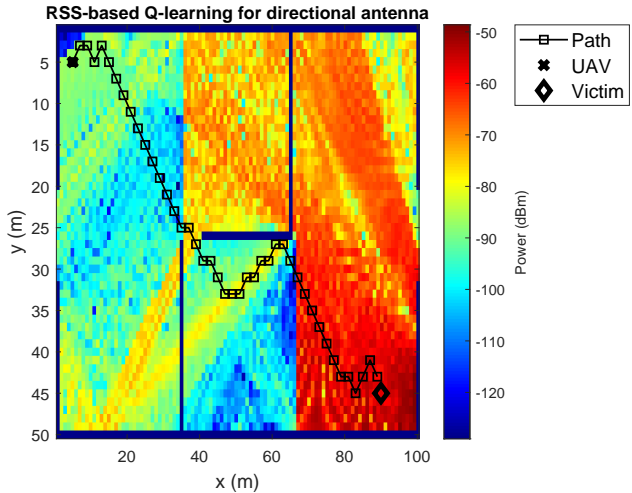}}
\caption{Trajectory followed by UAV on increasing iterations to 30000.}
%\label{fig:30000}
\end{figure}

\begin{table}[t]
\label{tab:results_30000_iter}
\caption{Results for 30000 Iterations on floor plan 1.}
\begin{center}
\begin{tabular}{cc}
\hline
Parameter & Value\\
\hline
UAV start position& (5~m,5~m) \\
Target location&	(45~m,90~m)\\
Number of iterations	&30000 \\
Total time & 45.10~s\\
Trajectory length& 111.34~m \\
Total steps taken	& 43 \\
\hline
\end{tabular}
\label{tab1:30000}
\end{center}
\end{table}

It can be observed that the trajectory followed by the UAV remains the same. As the number of iterations increase, the time elapsed also increases. However, this is done at the cost of precise and accurate results.

The other parameter which we have taken into consideration for comparison was antenna type. The results of the SAR operation using a directional antenna were compared to that of an omnidirectional antenna. The following are the results obtained on using an omnidirectional antenna for 30000 iterations on floor plan 1.

\begin{figure}[t]
\centering{\includegraphics[width=9cm]{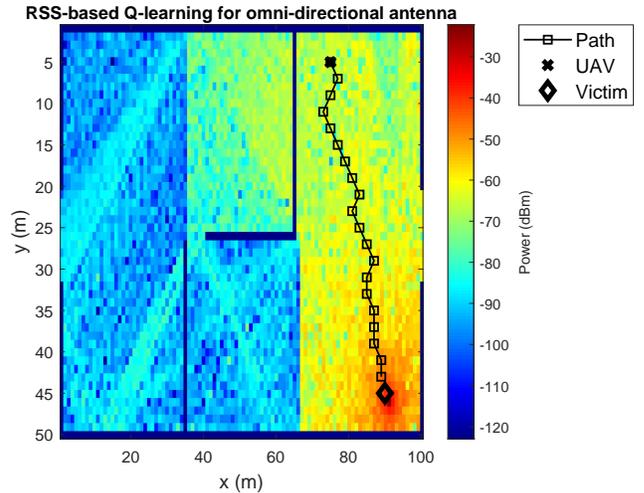}}
\caption{Results obtained by the use of an omnidirectional antenna by UAV for floor plan 1.}
\label{fig:omnidirectional1}
\end{figure}

\begin{figure}[t]
\centering{\includegraphics[width=9cm]{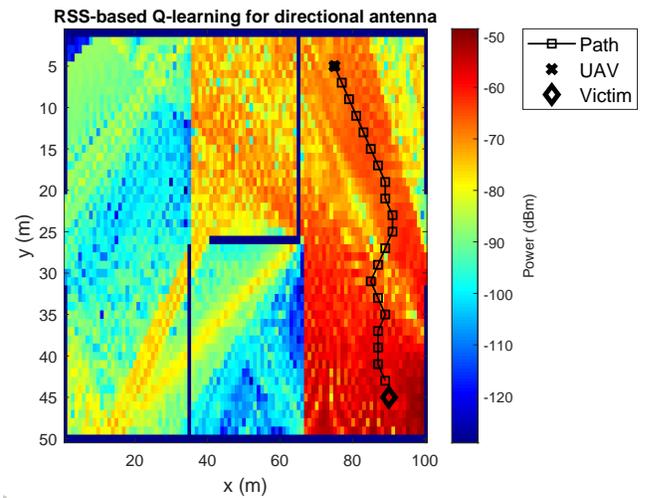}}
\caption{Steps taken by UAV using directional antenna on floor plan 1.}
\label{fig:directional1}
\end{figure}

\begin{table}[t]
\caption{Comparison of Results for Directional and Omnidirectional Antennas.}
\begin{center}
\begin{tabular}{ccc}
\hline
Parameters & Directional & Omnidirectional \\
\hline
UAV start position& (45~m, 90~m) &	(45~m, 90~m)\\
Target location	& (5~m, 7~m) & (5~m, 7~m) \\
Number of iterations&	30000&	30000 \\
Total Time &	16.56~s & 17.10~s \\
Trajectory	length& 50.43~m	& 50.43~m \\
Total steps taken & 20 & 20\\
\hline
\end{tabular}
\label{tab1:comparison}
\end{center}
\end{table}

\begin{figure}[t]
\centering{\includegraphics[width=7.5cm]{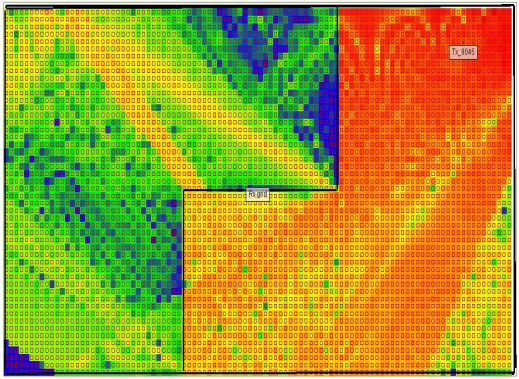}}
\caption{RSS heat map for 2.4~GHz transmit signal.}
\label{fig:wireless1}
\end{figure}

\begin{figure}[t]
\centering{\includegraphics[width=7.5cm]{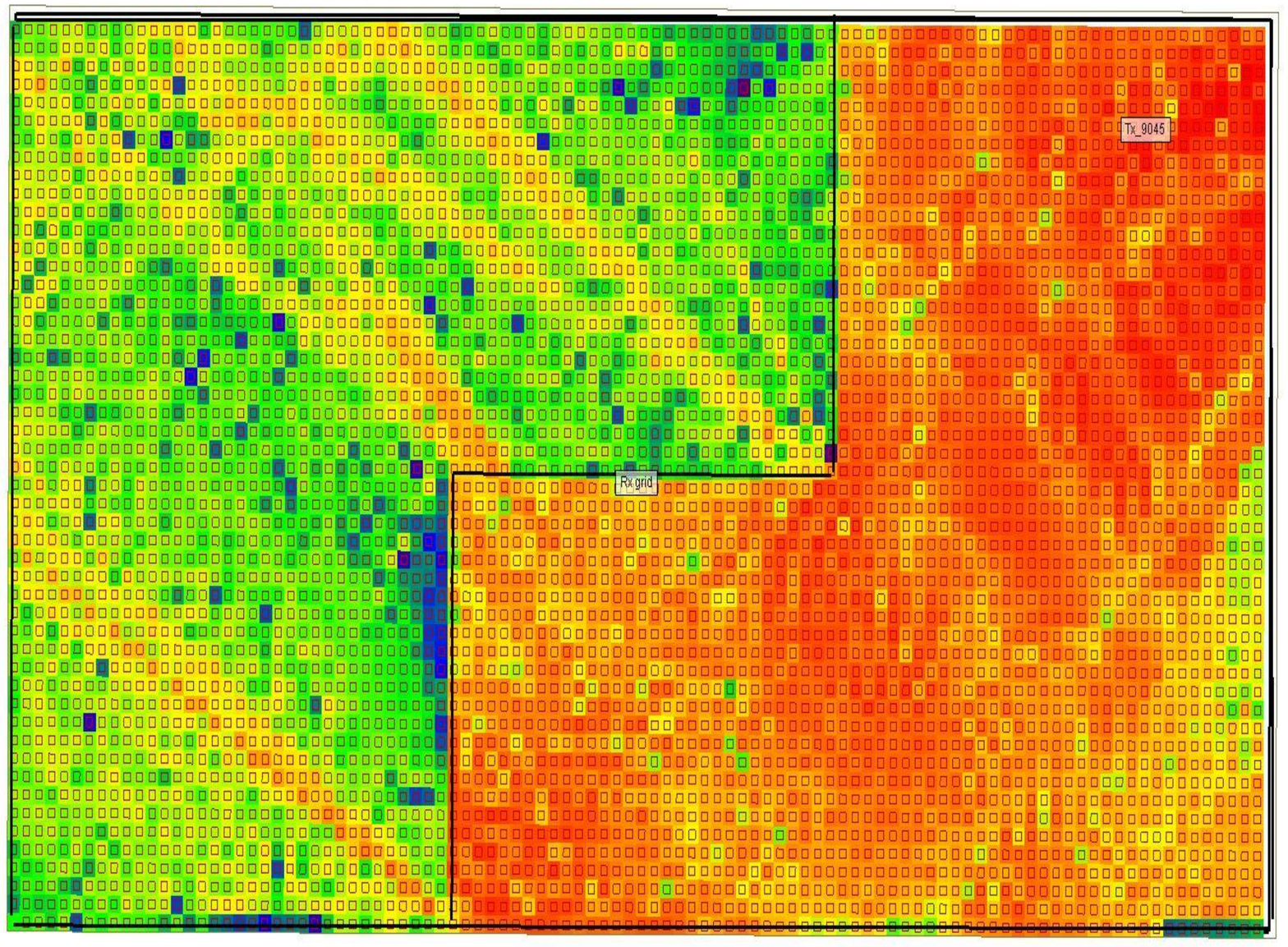}}
\caption{RSS heat map for 5~GHz transmit signal.}
\label{fig:wireless2}
\end{figure}

From Table~\ref{tab1:comparison}, and Figs.~\ref{fig:omnidirectional1}-\ref{fig:wireless2}, it can be inferred that if the UAV’s start location and the victim’s position was static, then the time to converge for a directional antenna is lower than that for the omnidirectional antenna. From this, we can infer that a directional antenna proves to be more efficient for the case of time dependent SAR operation. The realization of an omnidirectional antenna is possible only for simulation purpose while, in real time a directional antenna with low feedback loops, isolation and higher directivity is difficult to design. From Fig.~\ref{fig:omnidirectional1} and Fig.~\ref{fig:directional1}, we observed that the received signal strength for an omnidirectional and directional antenna is different for the same obstacles and indoor environment. We can thus deduce that a directional antenna is efficient even in environments having weak signals.

We have also compared the performance of the SAR operation in different frequency bands. Radio frequency waves at lower frequencies propagate further than radio frequency waves at higher frequencies. Thus, from Fig.~\ref{fig:wireless1} and Fig.~\ref{fig:wireless2}, we observe that the RSS of 2.4~GHz band is better in comparison to 5~GHz band. Additionally, we can infer that the longer wavelength of 2.4~GHz RF versus 5~GHz RF means that the 2.4~GHz signal will propagate through typical construction walls to a greater degree than the 5~GHz signal~\cite{rf}. 
From the results mentioned above, we can broadly infer that as the area of the indoor environment increases, the UAV takes greater time to reach the target. We cannot make this claim concrete as the time taken by the UAV is also dependent on factors like the obstacles in the indoor environment, received signal strength and the net distance between the UAV and the victim.
%\par There can be several future directions where this work can be extended, by relaxing the assumptions in this work.  Selective parameters of the search and rescue operation can be altered for real time realization.

\section{Conclusion}
In this paper, we studied a SAR operation by sensing the RSS of the RF signals emitted by the victim’s smart-device in a GPS-denied indoor environment. We considered that the UAV and the victim's smart device are equipped with directional antennas. We modeled two different indoor scenarios in commercial ray-tracing software and extracted the RSS values which were used to define the states and the rewards for our envisioned Q-learning model. We also carried out simulations for two different frequencies, 2.4~GHz and 5~GHz. Our results show that the change in frequency band resulted in different propagation patterns which affected the convergence time of the UAV. The convergence time required by the directional antenna was turned out to be lesser than that of the omnidirectional antenna. The change in convergence time was due to the concentration of transmitted signals in a specific direction in the case of the directional antenna. Therefore, the use of directional antenna as opposed to an omnidirectional antenna used in~\cite{chowdhury2019rssbased} proves to be a better choice in weak signal environments. 
%par In conclusion, we have leveraged the model free RL technique which uses an $\epsilon$ -greedy approach to find victims. The UAV is equipped with a configurable directional antenna to rescue targets within a shortest possible time interval.
\par Probable improvements can be done by considering real-time instances that need SAR operations. One of the possible alterations can be done by allowing the UAV to transmit RF signals in collaboration with a transmitting victim device. This will lead to the generation of RSS at each UAV location dynamically. This will overcome the limitation of the static victim position. Implementation of a multi-agent, multi victim scenario is another modification. The addition of multiple agents and victims in the environment will replicate a real-time scenario adding faster rescue of victims. For implementing this, the interference caused due to the transmission of the RF signals from multiple victims at the same time needs to be considered. Aspects of UAV battery-life, speed, and 3D navigation can also be further explored for more realistic analysis.

\bibliographystyle{IEEEtran}
\bibliography{references}

%\bibliography{ref.bib}
\end{document}